%% file: thintreevalidation.tex
\input{preamb}

\begin{document}

\maketitle

\begin{abstract}
	An \emph{$\alpha$-thin tree} $T$ of a graph $G$ is a spanning tree such that every cut of $G$ has at most an $\alpha$ proportion of its edges in $T$. The \emph{thin tree conjecture} proposes that there exists a function $f$ such that for any $\alpha > 0$, every $f(\alpha)$-edge-connected graph has an $\alpha$-thin tree. Aside from its independent interest, an algorithm which could efficiently construct an $O(1)/k$-thin tree for a given $k$-edge-connected graph would directly lead to an $O(1)$-approximation algorithm for the asymmetric travelling salesman problem (ATSP)~\cite{Gharan2011}. However, it was not even known whether it is possible to efficiently \emph{verify} that a given tree is $\alpha$-thin. We prove that determining the thinness of a tree is $\coNP$-hard.
\end{abstract}

\section{Introduction}\label{sec:intro}

Let $G$ be an undirected and unweighted graph. We denote by $V(G)$ the set of vertices in $G$, and by $E(G)$ the set of edges. Given a set $A \subseteq V(G)$, we denote by $\delta(A)$ or $\delta_G(A)$ the set of edges of $G$ with exactly one endpoint in $A$.

\begin{definition}
	A subgraph $S$ of a graph $G$ is \emph{$\alpha$-thin} if it contains at most an $\alpha$ proportion of the edges of each cut; that is, for every cut $C = (A, B)$ of $G$ (with $A \subseteq V(G)$, $B = V(G)\setminus A$):

	\[|\delta_S(A)|/|\delta_G(A)| \leq \alpha\]
\end{definition}

The following conjecture was proposed by Goddyn~\cite{Goddyn2004}:

\begin{conjecture}[Thin Tree Conjecture]
	There exists a function $f$ such that, for any $\alpha>0$, every $f(\alpha)$-edge-connected graph has an $\alpha$-thin spanning tree.
\end{conjecture}

In addition to a variety of other implications~\cite{Klein2023}, if it were possible to construct an $O(1)/k$-thin tree given a $k$-edge-connected graph, this would directly yield an $O(1)$-approximation algorithm for both the standard asymmetric traveling salesman problem (ATSP)~\cite{Gharan2011} and a ``bottleneck'' version of the ATSP (wherein the objective to minimize is the edge traversed of greatest weight)~\cite{An2021}. The former has already been produced through other methods, but the latter would be an improvement on the best known algorithm~\cite{An2021}. Constant-factor thin trees are known to be constructible in certain classes of graph~\cite{Gharan2011}, but in general graphs $\alpha$-thin trees are not known to exist for any $\alpha$ which is not a function of $n = V(G)$.

In fact, as far as we are aware, there has been no published result determining the difficulty of \emph{verifying} a given spanning tree as $\alpha$-thin. Variously this problem has been described as not being known to have a polynomially-checkable certificate~\cite{Klein2023} or as being $\NP$-hard (without proof)~\cite{Alghasi2023}. We present a proof that the problem of verifying whether a spanning tree is $\alpha$-thin, stated below, is $\coNP$-complete.

\begin{problem*}[Thin Tree Verification Problem (\TTVtext)]\label{prob:TTV}
	\leavevmode\vbox{}\\
	\textbf{Instance:} A graph $G$, a spanning tree $T$ of $G$, and a value $\alpha \in [0,1]$.\\
	\textbf{Output:} (YES) if $T$ is $\alpha$-thin---that is, if for all partitions $A$ of $G$,
	\[|\delta_T(A)|/|\delta_G(A)| \leq \alpha\]
	---and (NO) otherwise.
\end{problem*}

\begin{theorem}[restate=thmmain,
		name=Hardness of Thin Tree Verification]\label{thm:main}
	{\ThinTreeValid} is $\coNP$-complete.
\end{theorem}

\begin{remark*}
	The proof presented does not offer any guarantees of the approximation hardness of {\TTVtext}, nor does it imply an obvious way to approximate {\TTVtext}; it is thus of limited value either in demonstrating directly that a spanning tree within a constant factor of the edge connectedness of a graph cannot be found or in finding such a spanning tree. We present this work in hopes that it may be useful indirectly.
\end{remark*}

\section{Proof Strategy}\label{sec:prelim}

The proof of $\coNP$-hardness will consist of a reduction in turn through four problems, initially from the unweighted version of the classical problem {\MCtext}:

\begin{problem*}[Unweighted Maximum Cut Problem, {\MCtext}]\label{prob:maxcut}
	\leavevmode\vbox{}\\
	\textbf{Instance:} An (undirected, unweighted) connected graph $G$ and an integer $k$.\\
	\textbf{Output:} (YES) if there exists a partition $(A,B)$ of $G$ such that $|\delta(A)| \geq k$, and (NO) otherwise.
\end{problem*}

{\MCtext} is well-known to be $\NP$-hard~\cite{Garey1974}.

First, we reduce to the problem of finding the cut with the maximal \emph{average} weight amongst all cuts:

\begin{problem*}[Maximum Average Cut Optimisation Problem, {\MACOtext}]\label{prob:MACO}
	\leavevmode\vbox{}\\
	\textbf{Instance:} A connected graph $G$ with weights $\mathbf{w}$ such that $\forall w_e \in \mathbf{w}, w_e \in \{-1, 1\}$.\\
	\textbf{Output:} A nontrivial partition $(A,B)$ of $G$ such that, for all nontrivial partitions $A',B'$ of $G$,

	\[\left(\sum_{e\in\delta(A)}w_e\right)/|\delta(A)| \geq \left(\sum_{e\in\delta(A')}w_e\right)/|\delta(A')|\]
\end{problem*}

Next, we reduce {\MACOtext} to a natural decision variant (on multigraphs, for ease of proof):

\begin{problem*}[Maximum Average Cut Decision Problem, {\MACtext}]\label{prob:MAC}
	\leavevmode\\
	\textbf{Instance:} A connected multigraph $G$ with weights $\mathbf{w}$ such that $\forall w_e \in \mathbf{w}, w_e \in \{-1, 1\}$, and an integer $k \geq -1$.\\
	\textbf{Output:} (YES) if there exists a nontrivial partition $(A,B)$ of $G$ such that

	\[\left(\sum_{e\in\delta(A)}w_e\right)/|\delta(A)| \geq k,\]

	and (NO) otherwise.
\end{problem*}

Finally, we reduce this decision problem to the complement of {\ThinTreeValid}:

\begin{problem*}[The Complement of the Thin Tree Validation Problem, {\TTVCtext}]\label{prob:TTVC}
	\leavevmode\\
	\textbf{Instance:} A connected graph $G$, a spanning tree $T$ of $G$, and a value $\alpha \in [0,1]$.\\
	\textbf{Output:} (NO) if $T$ is $\alpha$-thin, and (YES) otherwise.
\end{problem*}

Additionally, for any graph $G$ and subgraph $T$ of $G$, we define the \emph{$T$-thickness} of a cut $(A,B)$ as the proportion of the edges in $\delta(A)$ which are in $T$; that is, $|\delta_T(A)|/|\delta_G(A)|$. Note that a subgraph $T$ of $G$ is $\alpha$-thin if and only if no cut of $G$ has a $T$-thickness strictly greater than $\alpha$.

\section{Proof}\label{sec:main}
Observe that {\ThinTreeValid} is in $\coNP$; given a {\TTVtext} instance $(G, T, \alpha)$, any cut of $G$ with $T$-thickness greater than $\alpha$ certifies that it is a (NO) instance, and such a cut must exist. It thus remains to prove that it is $\coNP$-hard.

\begin{theorem}\label{thm:MACOtoMC}
	$\MaxCut \leq_T^P \MaxAvgCutOpt$.
\end{theorem}
\begin{proof}
	We will use the following algorithm, where $A \oplus B$ denotes the symmetric difference of $A$ and $B$:

	\begin{algorithm}[H]
		\SetKwFunction{MACO}{maco}
		\Input{An (undirected) connected graph $G$;\\
			an integer $k$;\\
			a subroutine \MACO{$G'$, $\mathbf{w}$} which solves {\MACOtext}.}
		\Output{${\MCtext}(G, k)$.}
		$C \leftarrow \emptyset$\;
		\nl\Repeat{$\displaystyle \sum_{e \in C'}w_e \leq 0$}{\label{algMC:loop}
			\For{$e \in G$}{
				$w_e \leftarrow
					\begin{cases}
						-1 & \quad \text{if }e\in C    \\
						1  & \quad \text{if }e\notin C
					\end{cases}
				$\;
			}
			$C' \leftarrow \delta($\MACO{$G$, $\mathbf{w}$}$)$\;
			$C_{\text{prev}} \leftarrow C$\;
			$C \leftarrow C \oplus C'$\;
		}
		Output (YES) if $|C_{\text{prev}}| \geq k$; otherwise, output (NO).\;
		\caption{Solving {\MCtext} using {\MACOtext}.}\label{algMC}
	\end{algorithm}

	Note that at each iteration of the loop at \autoref{algMC:loop} except the last, $C'$ contains more edges which are not shared with $C$ than it contains edges shared with $C$. This is because the average weight of a cut in $G$ with weights $\mathbf{w}$ is positive if and only if there are more edges in the cut outside $C$ than in it; each edge outside adds $1$ to the numerator, and each edge inside adds $-1$. Every iteration, then, will increase the number of edges in $C$ until the first iteration with nonpositive average weight, at which point the loop at \autoref{algMC:loop} will terminate.

	Thus, for each iteration except the last, the size of $C$ is strictly increasing; since $|C| \leq |E(G)|$, there will be at most $|V(G)|^2$ iterations. Since each iteration completes with linear runtime and has one call to the {\MACOtext} algorithm, \autoref{algMC} calls the {\MACOtext} subroutine a polynomially bounded number of times with a polynomially bounded amount of extra runtime.

	The space of cuts of a graph is closed under the symmetric difference, so each $C$ produced is in fact a cut of $G$. Further, we have that if there exists a cut $C^*$ of $G$ such that $|C^*| > |C|$, then $C^* \oplus C$ will have positive value in the {\MACOtext} instance $(G, \mathbf{w})$. Hence the {\MACOtext} subroutine will not return a cut of nonpositive value, and the algorithm will not terminate; it will only terminate when a cut of maximal size in $G$ is obtained. So, given that $C_{\text{prev}}$ is a cut of maximum size in $G$ at the point of termination, the algorithm will output (YES) if there exists a cut in $G$ with size at least $k$, and (NO) otherwise.
\end{proof}

\begin{theorem}\label{thm:MACtoMACO}
	$\MaxAvgCutOpt \leq_T^P \MaxAvgCut$.
\end{theorem}
\begin{proof}
	We use \autoref{algMACO}.

	\begin{algorithm}
		\SetKwFunction{MAC}{mac}
		\Input{An (undirected) connected graph $G$;\\
			a set of (integer) weights $\mathbf{w}$ of $G$ such that $\forall w_e \in \mathbf{w}, w_e \in \{-1, 1\}$;\\
			a subroutine \MAC{$G'$, $\mathbf{w}'$, $k$} which solves {\MACtext}.}
		\Output{A cut of maximal average size in $(G, \mathbf{w})$.}
		\nl{}Set $k$ to the maximum average weight of a cut in $(G, \mathbf{w})$, found using an exhaustive search with \MAC{$G$, $\mathbf{w}$, $k$} over all possible maximum average weights (i.e., for each possible weight, run \MAC{$G$, $\mathbf{w}$, $k$}, and take the largest for which ${\MACtext}(G,\mathbf{w}, k)$ is (YES)).\label{algMACO:search}\;
		$G' \leftarrow G$\;
		\nl\While{$|V(G')| > 2$}{\label{algMACO:outloop}
			\nl\For{$(v, u) \in V(G')\times V(G')$}{\label{algMACO:inloop}
				$G'' \leftarrow$ the multigraph formed by contracting the vertices $v$ and $u$ in $G'$.\;
				\If{\MAC{$G''$, $\mathbf{w}$, $k$}}{
					$G' \leftarrow G''$\;
					break\;
				}
			}
		}
		Output the partition of $G$ given by the classes of vertices identified with each of the vertices of $G'$.
		\caption{Solving {\MACOtext} using {\MACtext}}\label{algMACO}
	\end{algorithm}

	First, note that \autoref{algMACO:search} can be completed in a polynomial number of checks to the {\MACtext} subroutine; all average weights are of the form $p/q$ with $|p| \leq q \leq n$ integral, where $n = V(G)$. Thus there are $O(n^2)$ weights to check. Further, the minimum possible average weight of any cut is $-1$, so all inputs are valid $\MACtext$ instances.

	We proceed to the loops. Since we contract a pair of vertices in each iteration of the loop at \autoref{algMACO:outloop}, and the loop at \autoref{algMACO:inloop} involves at most one iteration per edge of $G$, if this algorithm terminates it does so in polynomially many calls to the {\MACtext} subroutine and with polynomial extra time. To see that it does terminate, note that if there exists a maximum average cut which partitions the vertices $v$ and $u$ into different sets, contracting those vertices does not reduce the value of the maximum average cut of the graph; thus, in any graph with more than two vertices there exist at least two vertices which can be contracted while maintaining the size of the maximum average cut. Thus, the loop at \autoref{algMACO:inloop} always finishes with a pair of vertices contracted; eventually, $G'$ consists of two vertices.

	It is also clear that the cut in $G$ associated with $G'$ at the end has an average weight of $k$, since the only nontrivial cut in $G'$ has weight $k$, and the edges in that cut have the same weights as the equivalent edges of the associated cut in $G$. Thus, this algorithm produces a cut with maximal average weight.
\end{proof}

\begin{theorem}\label{thm:TTVCtoMAC}
	$\MaxAvgCut \leq^P_m \ThinTreeValidC$.
\end{theorem}
\begin{proof}
	We will show this via Karp reduction, producing from the triplet $(G, \mathbf{w}, k)$ a graph $G'$, a tree $T$, and a value $\alpha$ such that the {\TTVCtext} instance $(G', T, \alpha)$ is a (YES) instance if and only if the {\MACtext} instance $(G, \mathbf{w}, k)$ is a (YES) instance.

	As a brief sketch, we will construct a ``blown up'' graph which has a large clique for each vertex in the original graph and a collection of edges between those cliques for each edge in the original graph (see \autoref{fig:transform}, though note that the actual number of vertices and edges is not accurately represented). We will then construct a tree $T$ and a threshold $\alpha$ such that all cuts in the original graph with an average weight of at least $k$ correspond to cuts in the new graph which have more that an $\alpha$-fraction of edges in $T$ (violating the property that $T$ is $\alpha$-thin), and vice-versa. Intuitively, we arrange for the number of edges in $T$ between cliques to be proportional to the weight of the edge between the associated vertices in the original graph, and we ensure that the cliques are so large that the cuts which split a clique have an extremely low $T$-thickness (and so are irrelevant for determining what the thinness of the tree is, as it is determined by only the cut of greatest thickness). To simplify the analysis, we make the cliques larger than they need to be to make this property hold; the number of vertices we produce in each clique is $24\cdot|E(G)|$, but a much smaller number would suffice at the cost of complicating the proof.

	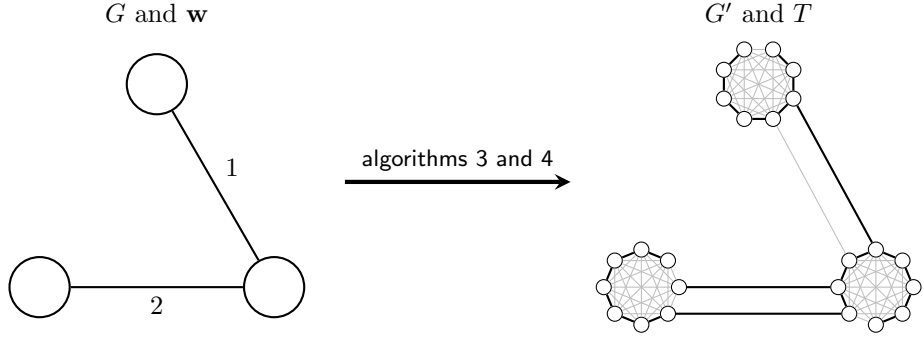
\begin{figure}
		\centering
		\begin{tikzpicture}
			\tikzset{simple_node/.style={circle, draw, thick, fill=white, minimum size=8mm}}
			\tikzset{clique_node/.style={circle, draw, thin, fill=white, inner sep=0pt, minimum size=2mm}}
			\tikzset{nontree_edge/.style={thin, gray!50}}

			\begin{scope}
				\node[anchor=south] at (0, 2.5) {$G$ and $\mathbf{w}$};
				\graph [clockwise=3, radius=1.8cm, phase=90, nodes={simple_node}, edges={thick}] {
				1[as=];
				2[as=];
				3[as=];
				1 --["1"] 2 --["2"] 3;
				};
			\end{scope}

			\draw[->, >=stealth, ultra thick] (2.5, 0.5) -- (5.5, 0.5)
			node[midway, above, font=\small\sffamily] {algorithms 3 and 4};

			\begin{scope}[xshift=8cm]
				\node[anchor=south] at (0, 2.5) {$G'$ and $T$};
				\graph [clockwise=3, radius=1.8cm, phase=90, nodes={clique_node}, edges={thick}] {
				c1[coordinate];
				c2[coordinate];
				c3[coordinate];

				subgraph K_n [n=8, name=C1, clockwise, radius=5mm, phase=67.5, nodes={clique_node, shift={(c1)}, as=}, edges={nontree_edge}];

				subgraph K_n [n=8, name=C2, clockwise, radius=5mm, phase=225, nodes={clique_node, shift={(c2)}, as=}, edges={nontree_edge}];

				subgraph K_n [n=8, name=C3, clockwise, radius=5mm, phase=0, nodes={clique_node, shift={(c3)}, as=}, edges={nontree_edge}];

				C1 3 -- C2 4;
				C1 4 --[nontree_edge] C2 3;
				C2 1 -- C3 2;
				C2 2 -- C3 1;

				\foreach \x [evaluate=\x as \xp using {int(\x+1)}] in {1,...,7} {
						C1 \x -- C1 \xp;
						C2 \x -- C2 \xp;
					};
				\foreach \x [evaluate=\x as \xp using {int(\x + 1)}] in {2,...,7} {
						C3 \x -- C3 \xp;
					};
				};
			\end{scope}
		\end{tikzpicture}
		\caption{A rough schematic of the transformation of $G$ and $\mathbf{w}$ into a ``blown up'' graph $G'$ and a tree $T$.}\label{fig:transform}
	\end{figure}

	\begin{algorithm}
		\Input{A connected multigraph $G$.}
		Let $m = |E(G)|$.\;
		$G' \leftarrow (\emptyset, \emptyset)$\;

		\For{$v \in V(G)$}{
			Add to $G'$ a clique $c_v$ consisting of $24m$ new vertices.\;
		}

		\For{$e = (v, w) \in E(G)$}{
			\nl{}Pick $3$ pairs $(v_0, w_0), (v_1, w_1), (v_2, w_2)$ of unique vertices in $G'$, one of each pair from $c_v$ and one of each from $c_w$, which do not yet have any incident edges to vertices outside of their respective cliques. (That is, pick pairs consisting of vertices that have not been chosen previously as part of a pair by this process.)\label{alg:Gcons:pairs}\;
			For each pair, add the edge $(v_i, w_i)$ to $G'$. Call this collection of three edges $d_e$.\;
		}
		\caption{Construction of $G'$}\label{alg:Gcons}
	\end{algorithm}
	\begin{algorithm}
		\Input{A connected multigraph $G$;\\
			a graph $G'$ constructed via \autoref{alg:Gcons} from $G$;\\
			a set of integral weights $\mathbf{w}$ on $G$ such that $\forall w_e \in \mathbf{w}, |w_e| \leq 1$.}
		Let $m = |E(G)|$.\;
		For each $v \in V(G)$, let $c_v$ be the associated clique in $G'$.\;
		For each $e \in E(G)$, let $d_e$ be the associated triplet of edges in $G'$.\;
		$S \leftarrow (\emptyset, V(G'))$\;
		\For{$v \in V(G)$}{
			Add a path to $T'$ connecting the vertices of $c_v$.\;
		}
		\For{$e = (v, w) \in E(G)$}{
			Add to $S$ $w_e + 2$ edges from among the edges in $d_e$.\;
		}
		$T \leftarrow S$\;
		\nl\While{$T$ is not a tree}{\label{alg:Tcons:cycles}
			Find a cycle $C$ in $T$.\;
			Pick an edge $e$ in $C$ which is internal to some clique of $G'$.\;
			Remove $e$ from $T$.\;
		}
		\caption{Construction of $T$}\label{alg:Tcons}
	\end{algorithm}
	Take the {\MACtext} instance $(G, \mathbf{w}, k)$. We first use \autoref{alg:Gcons} and \autoref{alg:Tcons} to construct an associated $G'$ and $T$.
	\needspace{4\baselineskip}
	\begin{claim}\label{claim:GTvalid}
		Both \autoref{alg:Gcons} and \autoref{alg:Tcons} can be performed in polynomial time and on any input matching their constraints.
	\end{claim}
	\begin{proof}
		It is easy to see that \autoref{alg:Gcons} and all parts of \autoref{alg:Tcons} save for the loop at \autoref{alg:Tcons:cycles} can be performed in polynomial time, and it is easy to see that all lines in both algorithms except for \autoref{alg:Gcons:pairs} of \autoref{alg:Gcons} can always be performed.

		In \autoref{alg:Gcons}, one can see that \autoref{alg:Gcons:pairs} is always possible by noting that any vertex has degree at most $m$, so at most $3m$ vertices of each clique must be selected by \autoref{alg:Gcons:pairs}; since $3m \geq 24m$ while $m \geq 1$, there are always enough vertices in each clique to select them in this manner.

		To see that the loop at \autoref{alg:Tcons:cycles} of \autoref{alg:Tcons} can be done in polynomial time, note that finding a cycle can be done in $O(m)$ time with any simple search algorithm, and that any cycle must necessarily contain a clique-internal edge (there exists no cycle in $G'$ consisting only of inter-clique edges, as no two inter-clique edges share a vertex). Since the number of edges in $T$ decreases by one for each iteration, this loop terminates in polynomial time.
	\end{proof}

	\begin{claim}\label{claim:Tspanning}
		The $T$ produced by \autoref{alg:Tcons} is a spanning tree of $G'$.
	\end{claim}
	\begin{proof}
		It is easy to see that the $S$ produced by \autoref{alg:Tcons} is a spanning subgraph of $G'$, since $G$ is connected and each clique of $G'$ is internally connected. Explicitly, one can construct a path $P'$ in $G'$ between any two vertices $v' \in c_v$ and $u' \in c_u$ by taking a path $P$ in $G$ from $v$ to $u$ and replacing each $e$ in $P$ with any edge in $d_e$, along with whichever edges in the clique-internal segments are necessary to complete the path. The produced $T$, then, must also be spanning, since removing a single edge from a cycle can never disconnect any vertices (each vertex in the cycle is still connected via the cycle, and no other vertex could be independently disconnected).
	\end{proof}

	Next, we pick an appropriate $\alpha$ to ensure that the {\MACtext} instance $(G, \mathbf{w}, k)$ is a (YES) instance if and only if the {\TTVCtext} instance $(G', T, \alpha)$ is a (YES) instance. Suppose that a cut $C' = (A', B')$ of $G'$ respects all clique-internal edges; that is, $A' = \bigcup_{i=0}^{k}c_{a_i}$ for some vertices $a_0, \dots, a_k \in G$. Identify such a cut with a cut $C = (A,B)$ of $G$, with $A = \{a_0, \dots, a_k\}$.

	\begin{lemma}\label{lemma:alpha}
		There exists an $\epsilon$ such that with $\alpha = (2+k)/3-\epsilon$, any cut $C$ of $G$ with average weight at least $k$ corresponds to a cut $C'$ of $G'$ which has a $T$-thickness strictly greater than $\alpha$. Likewise, any cut $C'$ of $G'$ which respects all clique-internal edges and has a $T$-thickness strictly greater than $\alpha$ corresponds to a cut $C$ of $G$ with average weight at least $k$.
	\end{lemma}
	\begin{proof}
		Proof provided in \autoref{subsec:lemmas}.
	\end{proof}

	\autoref{lemma:alpha} implies that if the {\MaxAvgCut} instance $(G, \mathbf{w}, k)$ is a (YES) instance, then the {\ThinTreeValidC} instance $(G', T, \alpha)$ is a (YES) instance.

	\begin{lemma}[restate=lemmathickcuts]\label{lemma:thickcuts}
		With $T$ constructed as in \autoref{alg:Tcons} and $\alpha = 2/3-\epsilon$ as in \autoref{lemma:alpha}, any cut $C'$ of $G'$ which partitions the vertices of a clique into different sets has an $T$-thickness lower than $\alpha$.
	\end{lemma}
	\begin{proof}
		Proof provided in \autoref{subsec:lemmas}.
	\end{proof}

	\autoref{lemma:thickcuts} implies that if the {\TTVCtext} instance $(G', T, \alpha)$ is a (YES) instance, then there is a cut in $G'$ with $T$-thickness strictly greater than $\alpha$ which respects all clique-internal edges (there exists some cut which is appropriately $T$-thick, and no cut which splits a clique can be $T$-thick, so it must be a cut which does not). By \autoref{lemma:alpha}, this corresponds to a cut in $G$ with average weight at least $k$; thus, the {\MACtext} instance $(G, \mathbf{w}, k)$ is a (YES) instance.

	Thus, the {\MACtext} instance $(G,\mathbf{w},k)$ is a (YES) instance if and only if the {\TTVCtext} instance $(G',T,\alpha)$ is a (YES) instance, and since the mapping $(G, \mathbf{w}, k) \mapsto (G', T, \alpha)$ can be performed in polynomial time, $\MACtext \leq_m^P \TTVCtext$.
\end{proof}

\needspace{2\baselineskip}
\thmmain*
\begin{proof}
	We have by \autoref{thm:MACOtoMC}, \autoref{thm:MACtoMACO}, and \autoref{thm:TTVCtoMAC} that:

	\begin{align*}
		                  & \MCtext \leq_T^P \MACOtext \leq_T^P \MACtext \leq_m^P \TTVCtext \\
		\Rightarrow \quad & \MCtext \leq_T^P \TTVCtext
	\end{align*}

	Hence {\TTVCtext} is $\NP$-hard; its complement, {\TTVtext}, is thus $\coNP$-hard. Since {\TTVtext} is in $\coNP$, it is then $\coNP$-complete.
\end{proof}

\subsection{Proofs of Remaining Lemmas}\label{subsec:lemmas}
\begin{proof}[Proof of \autoref{lemma:alpha}]
	We want that cuts in $G$ with average weight $\geq k$ correspond to cuts between cliques in $G'$ with $T$-thickness $> \alpha$. A given cut $C = (A, B)$ in $G$ corresponds to a cut $C' = (A', B')$ in $G'$ with $3$ times the number of edges between the partitions, and the sum of all $w_e$ with $e \in \delta(A)$ is equal to the number of edges of $\delta(A')$ that are in $T$ minus $2$ for each edge of $\delta(A)$. Thus,

	\begin{align*}
		        & \frac{\sum_{e \in \delta(A)} w_e}{|\delta(A)|}  \\
		= \quad & \frac{|\delta_T(A')|-2|\delta(A)|}{|\delta(A)|} \\
		= \quad & \frac{|\delta_T(A')|}{|\delta_{G'}(A')|/3}-2
	\end{align*}

	So:

	\begin{align*}
		k                             & \leq \left(\sum_{e \in \delta(A)}w_e\right)/|\delta(A)| \\
		\Leftrightarrow \quad (k+2)/3 & \leq |\delta_T(A')|/|\delta_{G'}(A')|
	\end{align*}

	However, we want any cut in $G'$ with an average proportion in $T$ \emph{strictly greater than} $\alpha$ to correspond to a cut in $G$ with an average weight \emph{at least} $k$ ($\leq$ is the complement of $>$, not $\geq$). Our desired $\alpha$ is then $(k+2)/3-\epsilon$ for some $\epsilon$ sufficiently small that there are no cuts of $G'$ with $T$-thickness in the interval $\halfopen{(k+2)/3-\epsilon,(k+2)/3}$. In order to find an $\epsilon$ which works, note first that the $T$-thickness of any cut in $G'$ is of the form $p/q$ for integers $p, q \leq |E(G')|$. Thus, if we take the set

	\[S = \{(p,q) : 0 \leq p, q \leq |E(G')|, p/q < (k+2)/3\}\]

	and $\alpha = \max_{(p,q) \in S}(p/q)$, there cannot exist any cut with $T$-thickness between $\alpha$ and $(k+2)/3$. We set $\epsilon$ to whatever value ensures that $\alpha$ equals this quantity.

	This guarantees that any cut in $G'$ with $T$-thickness greater than $\alpha$ and which does not assign different vertices of the same clique to different partitions corresponds to a cut in $G$ with an average weight at least $k$; likewise, cuts of this kind with $T$-thickness at most $\alpha$ have an average weight less than $k$.
\end{proof}

\begin{proof}[Proof of \autoref{lemma:thickcuts}]
	Consider any cut $C' = (A', B')$ which splits a clique in $G'$ into two. We will show that $C'$ has a $T$-thickness strictly lower than $(k+2)/3 > 1/3$ by considering first the internal edges of a single split clique along with all inter-clique edges, and then all other clique-internal edges.

	Consider some clique it splits, $c_v$, and call the number of vertices in a side of the partition with the fewest vertices $a$ (on the other side, there are $24m-a$ vertices). There are $(a)(24m-a)$ clique-internal edges, and at most $2a$ of them are in $T$. Assume that every single inter-clique edge is in $T$ and there are the maximum number of them; this is $3m$ edges. Thus, the $T$-thickness of the portion of $C'$ which includes only the edges in $c_v$ and all inter-clique edges is at most:
	\begin{align}
		\frac{3m+2a}{3m + a(24m-a)} & \leq \frac{3m+2a}{a(24m-a)}                                    \\
		                            & \leq \frac{3m}{a(24m-a)} + \frac{2}{24m-a}\label{eq:splitline}
	\end{align}

	Since $a \leq 12m$, we can bound the right term of \autoref{eq:splitline}:

	\[\frac{2}{24m-a} \leq \frac{1}{6m}\]

	The left term is maximised when $a(24m-a)$ is minimised. This happens when $a=1$:

	\[\frac{3m}{a(24m-a)} \leq \frac{3m}{24m} = \frac{1}{8}\]

	So the $T$-thickness of this portion of $C$ is at most $\frac{1}{8}+\frac{1}{6m} < \frac{1}{3}$. Hence, a cut that splits a single clique has a $T$-thickness strictly less than $(k+2)/3$, and at most $\alpha = (k+2)/3 - \epsilon$ as given above. We now consider the contributions of the edges internal to other split cliques, if any exist; for any such clique, following the bounds above, at most a $1/8$ portion of the edges contributed to $C$ by that clique are in $T$. Thus, it cannot be the case that considering the edges from additional split cliques increases the $T$-thickness of $C$ above $\alpha$.
\end{proof}

\section*{Acknowledgements}

We would like to thank Elena Grigorescu for advice and guidance, Greg Rosenthal for extensive proofreading and copyediting, and Kaleb Ruscitti for feedback.

\printbibliography{}

\end{document}

%% file: preamb.tex
\documentclass{article} 
\usepackage{graphicx}

\usepackage{fullpage,amsmath,amsthm,amsfonts,amssymb,bbold,thm-restate,needspace}

\usepackage[ruled,vlined]{algorithm2e}
\DontPrintSemicolon{}
\SetKwInOut{Input}{Input}
\SetKwInOut{Output}{Output}
\SetKwProg{Fn}{Function}{ is}{end}

\usepackage{hyperref}
\usepackage[nameinlink]{cleveref}
\usepackage{appendix}
\usepackage[style=alphabetic]{biblatex}
\usepackage{tikz}
\usetikzlibrary{graphs}
\usetikzlibrary{graphs.standard}
\usetikzlibrary{quotes}

\newtheorem{conjecture}{Conjecture}
\newtheorem{lemma}{Lemma}
\newtheorem{claim}{Claim}
\newtheorem{theorem}{Theorem}
\newtheorem*{theorem*}{Theorem}

\newtheorem{definition}{Definition}
\newtheorem*{problem*}{Problem}

\newtheorem{remark*}{Remark}

\newcommand{\NP}{\mathsf{NP}}
\newcommand{\coNP}{\mathsf{coNP}}
\newcommand{\MCtext}{\textsc{MaxCut}}
\newcommand{\MaxCut}{\hyperref[prob:maxcut]{\MCtext}}
\newcommand{\MACOtext}{\textsc{MaxAvgCutOpt}}
\newcommand{\MaxAvgCutOpt}{\hyperref[prob:MACO]{\MACOtext}}
\newcommand{\MACtext}{\textsc{MaxAvgCut}}
\newcommand{\MaxAvgCut}{\hyperref[prob:MAC]{\MACtext}}
\newcommand{\TTVtext}{\textsc{ThinTreeValid}}
\newcommand{\ThinTreeValid}{\hyperref[prob:TTV]{\TTVtext}}
\newcommand{\TTVCtext}{\textsc{ThinTreeValid\textsuperscript{C}}}
\newcommand{\ThinTreeValidC}{\hyperref[prob:TTVC]{\TTVCtext}}

\title{Thin Tree Verification is coNP-Complete}

\date{December 31, 2025}

\author{
Alice Moayyedi \thanks{David R. Cheriton School of Computer Science, 
University of Waterloo, Canada. 
\url{anelosima@proton.me}}}

\newcommand\halfopen[2]{\ensuremath{(#1,#2]}}